\documentstyle[floats,10pt,epsf,twocolumn,prl,aps]{revtex}

\begin{document}

\draft

\title{Comment on ``Electrical transport in junctions between unconventional
superconductors: Application of the Green function formalism''}

\author{Yu.~S.~Barash $^{a}$ and
A.~A.~Svidzinsky $^{b}$
\\}
\address{
$a)$ P.N. Lebedev Physical Institute, Leninsky Prospect 53, Moscow 117924,
Russia \\
$b)$ Department of Physics, Stanford University, Stanford CA 94305--4060, USA\\
}
\date{\today}

\maketitle

\begin{abstract}
\noindent {\bf Effect of surface pair breaking, entirely 
neglected by M.~Samanta and S.~Datta
[~Phys.Rev. B {\bf 57}, 10972 (1998)~], is quite important in considering
surface (~or interface~) quasiparticle bound states and associated
characteristics of junctions involving unconventional superconductors. The
whole class of bound states with nonzero energy is simply omitted within the
framework of the approach, using uniform spatial profile of the order parameter
up to the interface. The contribution of these bound states (as well as midgap
states) to current-voltage characteristics of the SIS' tunnel junctions
were studied for the first time in our earlier article. Dependence of midgap
state contribution to the Josephson critical current upon crystal to interface
orientations is shown as well to be fairly sensitive to the effect of surface
pair breaking.
}
\end{abstract}





In a recent article\cite{s98}, M.~Samanta and S.~Datta considered theoretically
electrical
transport of junctions involving unconventional superconductors. In particular,
they discussed contributions to junction properties from midgap surface states,
that arise in $d$-wave superconductors due to sign change of the order
parameter. They pointed out that the effect of midgap states is most
prominent for weakly coupled junctions (tunneling limit), concentrating in this
respect mainly on the first order theory in transmission coefficient. In
evaluating electric current across the junction the authors\cite{s98} neglected
from the very beginning all the effects of surface pair breaking, considering
order parameters from both sides of the junction to be equal to their bulk
values up to the junction barrier plane. In this Comment we demonstrate that
the approximations\cite{s98} lead to incorrect results for unconventional
superconductors, since effects of surface pair breaking is of crucial
importance for the I-V characteristics of tunnel junctions, especially due to
the appearence of surface quasiparticle states with nonzero energy. Also we
point out that correct theory for current-voltage characteristics of tunnel
junctions involving anisotropically paired superconductors was developed for
the first time in\cite{bs97}.

In contrast to s-wave isotropic superconductors, d-wave superconductors are
known to be quite sensitive to any inhomogeneities (impurities, surfaces,
interfaces).  In particular, for many of crystal to surface orientations the
order parameter turns out to be essentially suppressed on the tunnel barrier
plane.  Several important experimental methods used for studying the
anisotropic structure of the order parameter, for example, tunneling
measurements are, in turn, fairly sensitive to the superconducting properties
close to the surface of the sample. The effects of anisotropic pairing on the
tunneling density of states (the local quasiparticle spectrum at the surface),
the Josephson and quasiparticle current of SIS' and SIN tunnel junctions were
theoretically studied by taking account of surface pair breaking and
quasiparticle surface bound states
in\cite{bs97,bgz95,buch195,buch295,bbr96,bbs97,fog}.

Our main assertion is that the whole class of surface quasiparticle states is
omitted in\cite{s98} due to the disregard the surface pair breaking there.
Spatial profile of the order parameter suppressed near the surface, can be
considered as effective potential well for quasiparticles, which suffer Andreev
reflection towards the surface in this case. Andreev reflection processes along
with the conventional reflection from the surface, can result in forming
quasiparticle bound states (Andreev bound states) localized near the surface
within the characteristic length roughly of order of the superconducting
coherence length. Quasiparticle surface bound states with nonzero energy are
present in the case of surface pair breaking and do not exist for a uniform
spatial profile of the order parameter.  Only mid-gap surface states, having
supersymmetric origin, still exist for that uniform model profile. Thus, for
the order parameter, which is independent of the spatial coordinate up to the
surface or interface, one can find in the tunneling limit only peak at zero
energy in the local density of states, while all nonzero peaks taking place in
the presence of surface pair breaking turn out to disappear in the model.

Bound states with nonzero energy result in the anomalies of current-voltage
characteristics in the presence of externally applied voltage, described
in\cite{bs97,bbs97} and entirely omitted in\cite{s98}.  Positions and
characteristics of those anomalies turn out to be associated with positions and
types of extremal points of momentum dependence (dispersion) of bound state
energies.  Possible characteristic points on the I-V curves for SIS' and SIN
tunnel junctions are described in\cite{bs97,bbs97}, by using Eilenberger's
equations for the quasiclassical Green's functions combined with corresponding
boundary conditions and the microscopic expression for the tunnel current.

Midgap states are dispersionless bound states, so that their contributions to
junction characteristics can differ from the ones with nonzero energy.
In particular, if midgap states take place only from one side of a SIS' tunnel
junction, they contribute to the specific features of quasiparticle current
(and not to ac Josephson effect) at the voltage values, determined by anomalies
of the local density of states on the other bank of the junction. For example,
in the case of an isotropic $s$-wave superconductor on the other bank, with an
order parameter $\Delta$, quasiparticle current has peaks at voltage values
$eV=\pm \Delta$.  The peaks have inverse square-root ``divergent'' behavior in
the vicinity of $|eV|=|\Delta|$:  $I_{qp}\propto
\Theta(|eV|-|\Delta|)/\sqrt{|eV|^2-|\Delta|^2}$.  This particular result,
obtained for the first time in\cite{bs97}, was rederived in\cite{s98} in
disregarding the surface pair breaking.  There are various reasons for the
modification of this inverse square root singularity, at least sufficiently
close to the point $|eV|=|\Delta|$, in particular, due to the surface roughness
and bulk impurities. We do not discuss in the Comment this kind of effects,
determining the hight of the peaks, since it is not discussed at all
in\cite{s98}. We note that even in the absence of any surface pair breaking
there are some additional peaks and jumps of the conductance of SIS' junction,
involving anisotropically paired superconductors. Positions of the specific
features of the conductance are determined by the extremal points of the sum of
order parameters from both sides of the junction (and for the difference as
well, although not for sufficiently low temperatures), taken for incoming and
transmitted quasiparticle momenta\cite{bs97}. These specific features are
omitted in\cite{s98} as well.

As it is known, for SIN tunnel junction of high quality involving a $d$-wave
superconductor, the zero-bias anomaly of the conductance takes place due to the
midgap states. Let the normal metal suffer the superconducting transition at
lower temperature into a $s$-wave isotropic superconducting state. Then,
according to the result of Ref.\ \onlinecite{bs97} just stated above, the
zero-bias conductance peak should split into two peaks lying at the applied
voltage $eV=\pm \Delta(T)$. This splitting increases along with $\Delta(T)$
with further decrease of temperature. The behavior of I-V curves of this type
for the tunnel junction between $Bi_2Sr_2CaCu_2O_8$ (hight-$T_c$
superconductor) and $Pb$ ($s$-wave superconductor) was recently  observed
experimentally in\cite{sinha98}.
There are several experimental results, which could be interpreted as due to
zero energy bound states\cite{sinha98,cov,alff1,alff2,alff3,walter98}. At the
same time no anomaly in the Josephson critical current due to midgap states is
observed to our knowledge.  Manifestations of bound states with nonzero
energies are not noticed experimentally up to now as well.  Possibly, this is
due to reasons resulting in brodening of Andreev bound states (~e.g. due to
interface roughness\cite{bbr96,bbs97}~).

In the presence of midgap states from both sides of tunnel junction, they cause
the specific changes both in the quasiparticle current and in ac Josephson
effect\cite{bs97}. The effect of finite transmission of the barrier plane
beyond the tunneling limit may result in the shift of midgap states to nonzero
values of energy (~though for particular case of ``mirror'' tunnel junctions
and no phase difference beteen superconductors there is no shift for the midgap
states due to this reason~). These ``former midgap states'' take place for the
uniform model as well, in contrast to other interface bound states.

Disregarding the surface pair breaking is the common feature of many articles,
which consider surface (interface) bound states in $d$-wave superconductors
both in studying  current-voltage
characteristics\cite{s98,hu94,tan95,tan196,hurd97,hurd98,hljw98} and dc Josephson
effect\cite{tan296,tan97,s97,rb98}. It is worth noting, that the effect of
surface pair breaking can be of importance not only for studying the
current-voltage characteristics but in considering the Josephson critical
current as well. In particular, in the presence of essential suppression of the
order parameter at the surface there are deviations of temperature dependence
of the Josephson critical current $I_c$ from conventional Ambegaokar-Baratoff
behavior both near $T_c$\cite{bgz95} and at low temperatures\cite{bbr96}. For
example, in the vicinity of $T_c$ the Josephson critical current turns out to
be proportional to $(T_c-T)^2$ for sufficiently small surface values of the
order parameters on both banks of the junction. It is proportional to
$(T_c-T)^{3/2}$ if the order parameter vanishes only from one side of the
tunnel barrier plane.  Conventional Ambegaokar-Baratoff behavior $I_c\propto
(T_c-T) $ holds near $T_c$ in the absence of any essential suppression of the
order parameters at the interface.

Since the low-temperature anomaly of the Josephson critical current is
associated with the effect of midgap states\cite{bbr96}, the influence of
surface pair breaking on the characteristics of midgap states should be
discussed in this context. The occurrence of the zero-energy peak in the
tunneling density of states is unaffected by the self-consistency of the order
parameter. However, disregarding surface pair breaking can result in essential
overestimating the weight of the peak and, as a consequence, the Josephson
critical current.  Moreover, since surface pair-breaking is sensitive to the
crystal to surface orientation, disregarding its effect results in qualitative
changes in dependences of the peak height and $I_c$ upon the misorientation
angles of superconductors from both sides of the junction. To illustrate the
above assertion we consider the midgap state contribution to the Josephson
critical current between two identical $d$-wave superconductors, as a function
of the crystal orientation of superconducting electrodes with respect to the
specularly reflecting tunnel barrier plane. For the sake of simplicity we
consider symmetric tunnel junction, when crystal axes have the same orientation
in both electrodes.


\bigskip
\centerline{\epsfxsize=0.45\textwidth\epsfysize=0.65\textwidth
\epsfbox{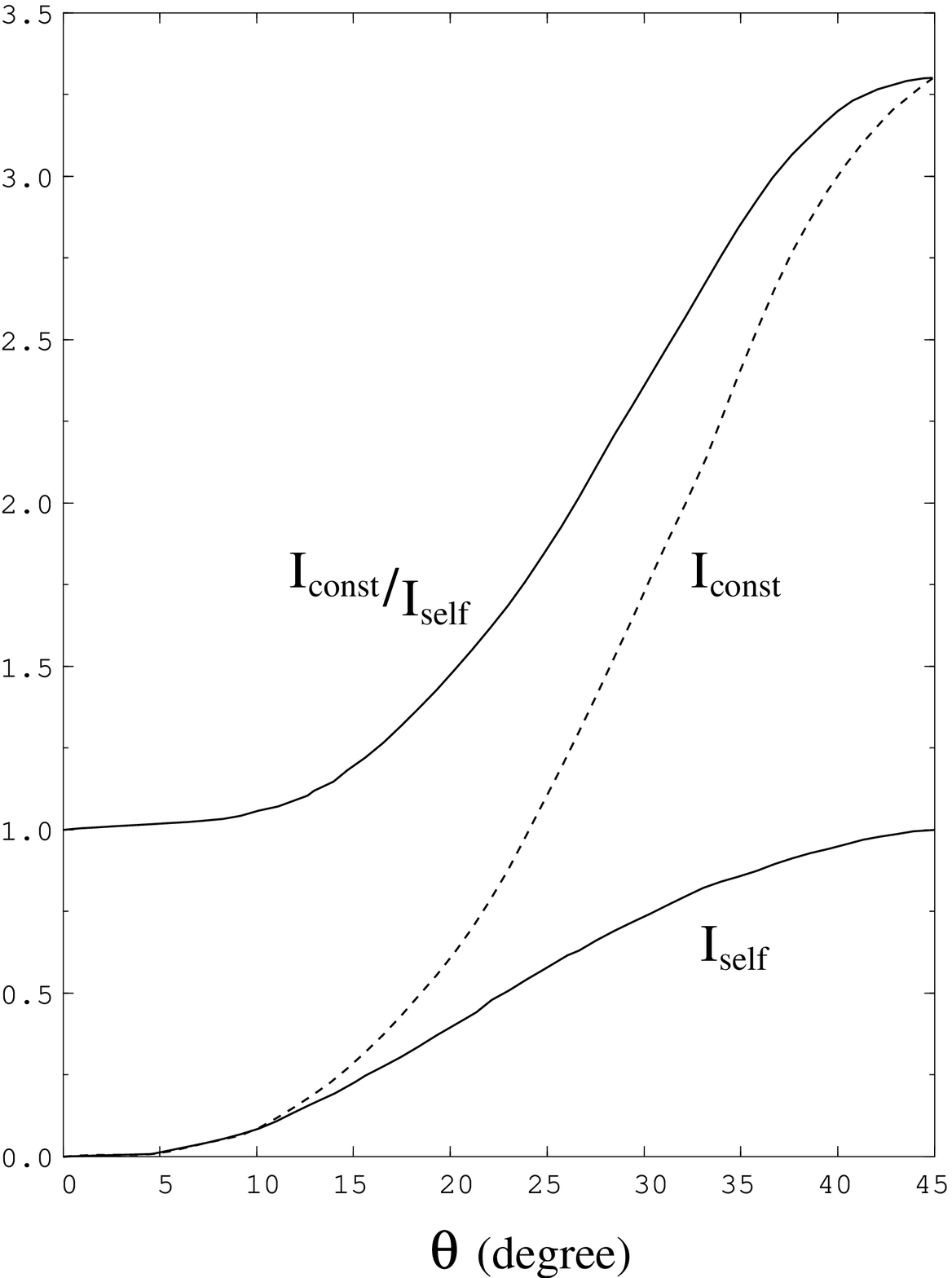}}

\vspace{0.3cm}

{\small  Fig.~1.  Midgap state contribution to the Josephson critical current
$I(\theta)$ as a function of misorientation angle $\theta$ at temperature
$T=0.1T_c$. Transmission coefficient $D$ is taken $\propto \cos^2\varphi$,
where angle $\varphi$ assigns the incoming momentum direction with respect to
the interface normal. Critical current $I_{const}(\theta)$ for the case of
constant order parameter (dashed line) is compared with calculations using a
self-consistently determined order parameter (lower solid line). Functions
$I_{self}(\theta)$ and $I_{const}(\theta)$ are normalized to the value
$I_{self}(\theta=45^o)$.  }
 
We use a cylindrical Fermi surface with an isotropic Fermi velocity and a
constant normal state density of states. The calculation of the spatially
dependent order parameter involves the pairing interaction, which we write as
$V(\bbox{p}_f,\bbox{p}_f^{\:\prime})=2V\cos(2\phi-2\theta)\cos(2\phi'-
2\theta)$. Here $\theta$ is the misorientation angle, that is the angle made by
the interface normal with the principal axis of the $d$-wave order parameter on
each side of the junction. In Fig.\ 1 we present the numerical results for the
midgap states contribution to the Josephson critical current (~which dominates
for sufficiently small temperatures~), as a function of the misorientation
angle. The lower solid line describes self-consistent results, while the dashed
line corresponds to neglecting the effect of pair breaking.  The quantities are
normalized to the self-consistent value of the Josephson critical current for
$\theta=45^0$.  The figure shows strong deviation of self-consistent angular
dependence from the non-selfconsistent one. This reflects the fact that
suppression of the order parameter near the tunnel barrier is essential for a
wide range of crystalline orientations and sensitive to change of the
orientation. The most significant deviation takes place for $\theta=45^0$, when
the order parameter is completely suppressed near the boundary.  For this
particular orientation disregarding the surface pair breaking results in
overestimation of midgap states contribution to the Josephson critical current more
than in three times. Similar difference can be seen in Fig. 2 in
Ref.\ \onlinecite{tk98}.
Our results demonstrate, in particular, the failure of the simple
orientation dependence $I_c(midgap)\propto\sin(2\theta_1) \sin(2\theta_2)$,
obtained in\cite{s97,s98} on the basis of a uniform model order parameter and
some additional approximations. One can show, however, that signs of
$I_c(midgap)$ and $\sin(2\theta_1) \sin(2\theta_2)$ coincide for the particular
pairing potential considered.

The authors of Ref.\ \onlinecite{s98} make also an accent on the fact that they
do not use quasiclassical approximation in their theory. Extension of the
theory beyond the quasiclassical approximation would be interesting both in the
case of any new qualitative features of the phenomena in question, or due to
noticeable quantitative corrections, containing powers of the parameter
$a/\xi$, where $a$ is the atomic scale. However, no new qualitative features of
nonquasiclassical nature are found in\cite{s98}. We believe, that taking
account of the effect of surface pair breaking within the quasiclassical
theory, is much more important from the qualitative point of view, while
quantitatively disregarding this effect makes the extension of the approach
beyond the quasiclassical approximation to be senseless.

Yu.S.B. acknowledges support by the Russian Foundation for Basic Research under
grant No.~96-02-16249. This work was supported in part by  Stanford University
(A.A.S.).

\end{document}